# Origami-inspired Cellular Metamaterial with Anisotropic Multi-stability

Soroush Kamrava,[a] Ranajay Ghosh,[b] Zhihao Wang,[a] and Ashkan Vaziri [a,*]

[a] Department of Mechanical and Industrial Engineering,
Northeastern University, Boston, MA 02115, USA
[b] Department of Mechanical and Aerospace Engineering
University of Central Florida, Orlando, FL 32816, USA

[*]Corresponding Author: vaziri@coe.neu.edu

**Abstract**:
Origami designs offer extreme reconfigurability due to hinge rotation and facet deformation. This can be exploited to make lightweight metamaterials with controlled deployability and tunable properties. Here, we create a family of origami-inspired cellular metamaterials which can be programmed to have various stability characteristics and mechanical responses in three independent orthogonal directions. The cellular metamaterials were constructed from their origami unit cell that can have one or two admissible closed-loop configurations. The presence of the second closed-loop configuration leads to the emergence of bi-stability in the cellular metamaterial. We show that the stability and reconfigurability of the origami unit cell, and thus the constructed cellular metamaterials, can be programmed by manipulating the characteristic angles inherited from the origami pattern. Two examples of such programmable metamaterial with bi-stability in out-of-plane direction and anisotropic multi-stability in orthogonal directions are presented. Our study provides a platform to design programmable three-dimensional metamaterials significantly broadening the application envelope of origami.

**Introduction:**

Geometry induced instabilities are ubiquitous in nature due to their importance in influencing large changes in mechanical response and adding extra functionality to the structure. Examples include the closure of Venus flytrap plant [1], trapping mechanism of the bladderworts [2], buckling of drying colloidal droplets [3], and osmotically shrinking polymeric capsules [4]. Including instability in a materials design can similarly lead to added functionality and rapid shape change. In this context, using origami structures to harness instability is an exciting new area of research. Twisted origami square [5], cylindrical origami with Kresling and Miura-ori pattern [6, 7], and rigid-foldable cellular structures with special hinge characteristics [8] are some examples of origami-inspired structures capable of exhibiting instability. More generally, over the past few decades, origami has truly evolved from an ancient Japanese art of paper folding into a rich scientific field bridging different disciplines [9]. However, in spite of the unlimited origami configurations possible, at a very fundamental level origami can be classified into two broad categories - *rigid-foldable* and *deformable* origami [10]. In the rigid-foldable origami, facets remain flat and the only source of deformation is rotation about the creases [11-14]. The assumption of rigid facets with zero thickness in the rigid-foldable origami causes the kinematics of origami to solely depend on the fold pattern and be independent of material and hinge properties [12, 15, 16]. These simplifications lead to straightforward correlations between the fold pattern and kinematics of origami folding [9, 14]. This very simplification also results in an inherently limited envelope of performance since for many applications, more degrees of reconfigurability may be desirable. This limitation can be overcome if more flexible facets are used resulting in deformable origami structures [16] which opens up new possibilities to develop bi-stable origami structures [5, 17].

Here, we propose a family of origami-inspired load bearing cellular structures with anisotropic programmable multi-stability. The multi-stability of the cellular structure originates from the bi-stability of individual deformable origami unit cells. Each unit cell structure is built by folding a Miura-ori strin g [18], which is a sequence of $n$ individual Miura-ori as shown in **Figure 1(A)**, to make a closed-loop configuration. Crease pattern of the Miura-ori string is defined by number of Miura-ori in the string ($n$), two repeating characteristic angles of $\alpha_1$ and $\alpha_2$ ($\alpha_1 > \alpha_2$) and the dimensions of $a$ and $H$, shown in **Figure 1(A)**. Angle $\theta$ is the dihedral angle between plates in flat and folded configurations, which varies from 0° to 90° and represents the level of folding. The Miura-ori string shown in this figure has $\alpha_1 = 77°$ and $\alpha_2 = 42°$ and folds from the flat configuration at $\theta = 0°$ and goes through a rigid foldable regime until it makes a closed-loop configuration at $\theta = 16°$. We investigate the formation of the closed-loop configuration for any arbitrary Miura-ori string with rigid facets based on two criteria. First, from the plane geometry of polygons, all closed-loop configurations should satisfy the internal angle condition $\eta_1 + \eta_2 = \pi(2 - 2/n)$, where $\eta_1$ and $\eta_2$ are the internal angles formed between longitudinal creases and shown in **Figure 1(B)** and $n$ is number of Miura-ori in the string (See Supporting Information for more details). The values of $\eta_1$ and $\eta_2$ at different folding levels are determined based on the governing equations of Miura-ori fold [8]:

$$\eta_1 = \pi - 2cos^{-1}(\frac{\cos \alpha_1}{\sqrt{1-\cos^2\theta \sin^2\alpha_1}}) \tag{1}$$

$$\eta_2 = \pi + 2cos^{-1}(\frac{\cos \alpha_2}{\sqrt{1-\cos^2\theta \sin^2\alpha_2}})$$

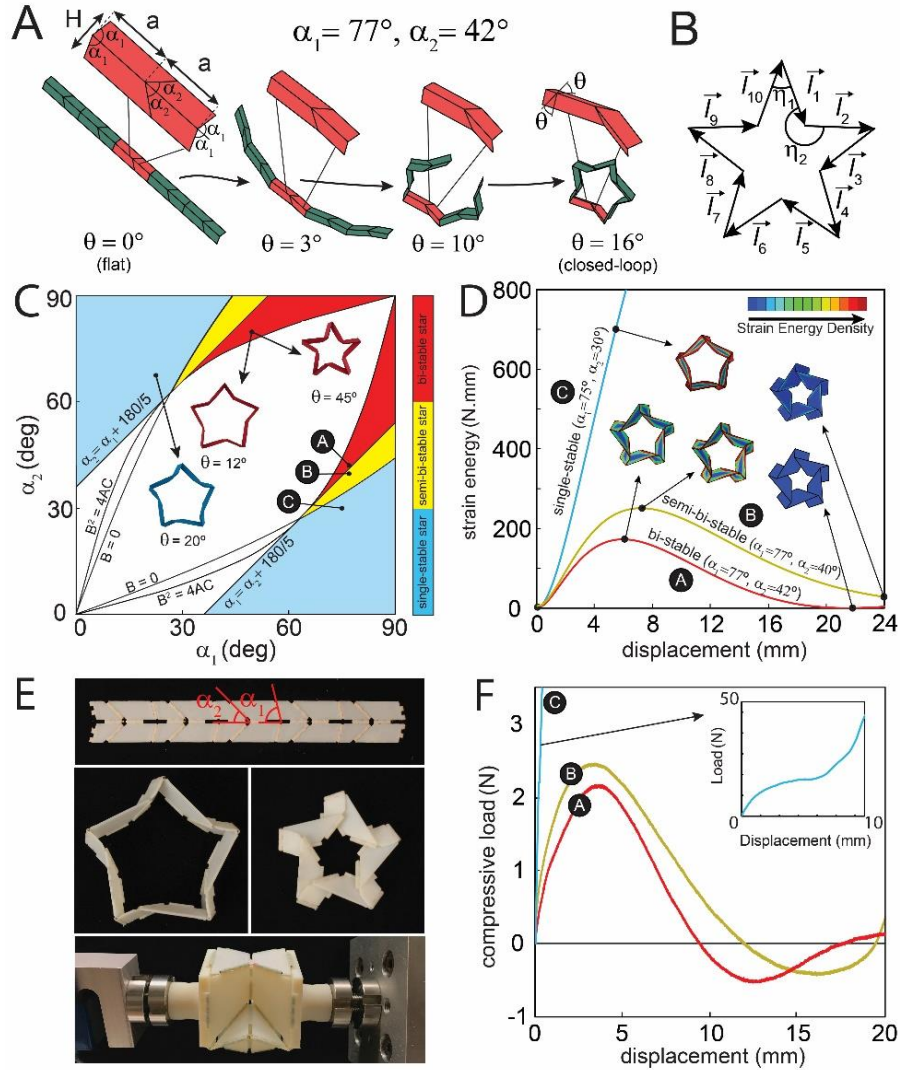

**Figure 1.** A) Folding an origami string consisted of five identical Miura-ori with characteristic angles of $\alpha_1$ and $\alpha_2$. The string folds from the flat configuration ($\theta = 0°$) to a closed-loop star-shaped configuration at $\theta = 16°$. B) The longitudinal creases of the closed-loop configurations that form internal angles of $\eta_1$ and $\eta_2$. C) Design map for stability analysis of star-shaped structures with different $\alpha_1$ and $\alpha_2$ angles. Blue, yellow, and red areas correspond to single-stable, semi-bi-stable, and bi-stable units, respectively. D) Numerically determined elastic strain energy versus applied compressive out-of-plane displacement for three different star-shaped structures denoted by $A$ ($\alpha_1 = 77°$, $\alpha_2 = 42°$), $B$ ($\alpha_1 = 77°$, $\alpha_2 = 40°$), and $C$ ($\alpha_1 = 75°$, $\alpha_2 = 30°$) showing the behavior of bi-stable, semi-bi-stable, and single-stable units, respectively. The distribution of strain energy density in each structure is shown at different compressive displacements. E) A 3D printed $n = 5$ origami string with $\alpha_1 = 77°$ and $\alpha_2 = 42°$ in the flat configuration which folds in to a five-pointed star-shaped structure. The resulting star-shaped structure is bi-stable with two stable configurations as shown. The structure can reversibly transform from configuration to another by applying a compressive/tensile loading. (F) Experimentally obtained load-displacement response of 3D printed samples $A$, $B$, and $C$.

Substituting **Equation 1** into $\eta_1 + \eta_2 = \pi(2 - 2/n)$ results in a trigonometric equation, which can be simplified to **Equation 2**.

$$Acos^4\theta + Bcos^2\theta + C = 0 \tag{2}$$

$$A = sin^2\alpha_1 sin^2\alpha_2$$

$$B = \frac{sin^2(\alpha_1 - \alpha_2)}{sin^2(\pi/n)} - (sin^2\alpha_1 + sin^2\alpha_2)$$

$$C = 1 - \frac{sin^2(\alpha_1 - \alpha_2)}{sin^2(\pi/n)}$$

Solving **Equation 2** gives us the values of $\theta$ in which a Miura-ori string described with $\alpha_1$, $\alpha_2$ and $n$ would make a closed-loop:

$$\theta_1 = cos^{-1}\left(\sqrt{\frac{-B + \sqrt{B^2 - 4AC}}{2A}}\right), \quad \theta_2 = cos^{-1}\left(\sqrt{\frac{-B - \sqrt{B^2 - 4AC}}{2A}}\right) \tag{3}$$

Note that in general **Equation 2** will have four roots. However, if we restrict our angles to $0° < \theta < 90°$, the two negative possibilities are eliminated. This ensures that for all practical purposes, only two possible type of solutions can emerge from **Equation 3.** Thus, any arbitrary Miura-ori string could have either zero, one or two closed-loop configurations which correspond to $\theta_1$ and $\theta_2$ both imaginary, one imaginary and one real and both real numbers, respectively. The second criterion requires two ends of Miura-ori string to meet each other at the closed-loop configurations which can be mathematically written as the summation condition $\sum_{i=1}^{2n} \vec{l_i} = 0$ where $\vec{l_i}$ is the $i^{th}$ middle crease vector shown in **Figure 1(B)** [19, 20].

Note that these possible configurations, which do not lead to facet bending, can be called zero energy configuration if hinge stiffness is neglected. These possible configurations can be plotted on a phase map. Such a phase map is shown in **Figure 1(C)** for Miura-ori strings with $n = 5$ and $\alpha_1$ and $\alpha_2$ ranging from $0°$ to $90°$. This phase map is symmetric with respect to the $\alpha_1 = \alpha_2$ line, which is expected from the geometry of the Miura-ori string. The white region is a forbidden zone where the given $\alpha_1$ and $\alpha_2$ would not lead to loop closure. The single solution configuration (i.e. unique loop configuration) is shown by the blue region and is called single-stability case. On the other hand, the red regions correspond to the folding pattern, which gives rise to two possible closed-loop solution and resulting in a bi-stable unit. However, although the single stable case is predicated on only one possible stable configuration there is a scenario that can yield another point of structural stability. We hypothesize that in this case, the second stable point will partially share the characteristic of the bi-stable case. In general, as soon as the origami is subjected to out-of-plane load, facet bending will make rigid origami conditions inapplicable. These would mean that both the internal angle and closed loop criteria will no longer be true. In other words, $\left|\sum \vec{l_i}\right| > 0$ and $|\eta_1 + \eta_2 - \pi(2 - 2/n)| > 0$. For a single stable unit this constraint will never be satisfied again. However, for the bi-stable unit this will be satisfied again at the second zero energy state. In other words, after an initial increase in $\left|\sum \vec{l_i}\right|$ and $|\eta_1 + \eta_2 - \pi(2 - 2/n)|$ they will begin to decrease. Hence, for another stable point to exist at a subsequent point, $d\left|\sum \vec{l_i}\right|/d\theta < 0$ and $d|\eta_1 + \eta_2 - \pi(2 - 2/n)|/d\theta < 0$. Whereas, for the bi-stable case, this will eventually lead to the zero strain energy minima, for the single stable case this manifests as the structure tries to return towards the zero-energy state. However, since it has only one possible stable configuration, the origami will stop in its track. Physically this would correspond to the configuration which leads to facet contact causing a sudden increase in stiffness of the structure after this point. We impose these restrictions on the single stable origami and find the emergence of another region called the semi-bistable region. This region is shown in yellow in **Figure 1(C).**

These arguments can be more readily seen through the strain energy landscape. We use finite element (FE) simulations on three origami samples $A$, $B$ and $C$ from the red, yellow, and blue regions using a commercially available FE code ABAQUS (Dassault Systemes). These origami units were modeled as a set of flat plates connecting to each other by ideal hinges with zero stiffness and zero friction to eliminate the effect of materials and merely investigate the geometry of units. Our simulations resulted in strain energy vs. out-of-plane displacement plots in **Figure 1(D).** For the single-stability case (sample $C$), the strain energy will continue to monotonically increase with load, as shown in **Figure 1(D)** using a blue line. For the bi-stable case (Sample $A$), an energy minimum is achieved once again with deformation at a later stage of folding level, as shown by the red line in **Figure 1(D).** The energy plot now reveals more clearly the nature of the second stability point for the semi-bi-stable case (sample $B$), shown using a yellow line. In this case a clear energy maximum is visible followed by a decline which would correspond to the negative derivative condition. But the decline is arrested by the face contact (the region after the face contact would be a very high energy state dictated by the contact configurations). Note that for this case, the deformation abruptly stops at the displacement of $24\ mm$ which corresponds to the contact between the facets. This strain energy configuration is not a global energy minimum (excluded from our mathematical relations) although it has lower energy than the neighboring configuration. The inset figures in **Figure 1(D)** show strain energy distribution in units $A$ and $B$ at the maximum and minimum levels of strain energy rather than initial configuration and also an arbitrary configuration of unit C. These curves can also be used to infer stiffness of the structures from the second derivative. Clearly, unit $C$ will not show any decrease in stiffness as its strain energy is monotonic with displacement. Unit $A$ and $B$ show clear inflections on their way between their stable configurations. Therefore, we can expect these structures to show an initial increase in stiffness, then decrease till they reach their second stable configuration. We can correlate these predictions using experiments. To this end we devise compression test on fabricated units corresponding to the geometry of $A$, $B$ and $C$ units described above. Fabrication of origami structure has been done in many ways in literature such as traditional paper folding [21] and using laser-cutting [22]. However, the elasticity of creases in these fabrication methods causes the deviation of real origami structures from the expected theoretical behavior. For example, the fabricated origami bellows failed to exhibit bi-stability, although their bi-stability is confirmed theoretically [19]. Here, we used 3D printed revolute hinges with zero energy at any angle to closely align with the assumption of zero-energy hinges in the theoretical and numerical modeling. All facets were fabricated using PolyJet 3D printing technique and connected to each other with brass pins allowing them to rotate freely (revolute hinge). **Figure 1(E)** shows a 3D printed sample of unit A ($\alpha_1 = 77°$ and $\alpha_2 = 42°$). Revolute hinges at both ends enable it to make a permanent closed-loop star-shaped unit. Also, both stable closed-loop configurations of this sample at $\theta_1 = 16°$ and $\theta_2 = 70°$ are shown in the figure. See the Supporting Movie for the transition of the structure between two stable configurations. The force-displacement curve of all three 3D printed $A$, $B$, and $C$ units under the out-of-plane compression test is measured using a ADMET testing machine positioned horizontally to eliminate the effect of structure weight. A picture of unit $A$ during the out-of-plane compression test is shown in **Figure 1(E)**. The force-displacement curve of units $A$, $B$, and $C$ are plotted in **Figure 1(F).** The force-displacement response of units $A$ and $B$ follow a positive, negative, and positive stiffness pattern and reaches the zero force at three distinct equilibrium configurations. These equilibria can be stable or unstable if they correspond to positive or negative force-displacement slope, respectively [23]. Hence, the first and third equilibriums are stable configurations which confirm the bi-stability of units $A$ and $B$. Also, force-displacement response of unit $C$ shows only one stable equilibrium configuration at the zero displacement.

The results presented in **Equations 2** and **3** were demonstrated for a string with five embedded Miura-ori which folds to a five-pointed star-shaped unit. However, these results can be used for all strings with $n$ Miura-ori which fold to make a $n$-pointed star-shaped unit. **Figure 2(A)** shows the prototypes of various $n$-pointed star-shaped units with $n$ ranging from 3 (minimum possible) to 10. The computational design maps of these stars are shown in **Figure 2(B)** in which the colored areas indicate the $\alpha_1$ and $\alpha_2$ values of the strings which would give rise to either a bi-stable or semi-bi-stable unit. As $n$ increases, the colored area moves toward the $\alpha_1 = \alpha_2$ line decreasing the possible multiple stability configuration. On the other extreme, when $n$ gets closer to three, the bi-stable region is possible in an increasingly thin slice of angles which restrict $\alpha_1$ to be near 90°. This angular configuration is near to the so called singular design point for the Miura-ori pattern and makes the unit dramatically sensitive to the fabrication errors [18]. To further study the characteristics of different star-shaped units, we modeled the eight units shown in **Figure 2(A)** and performed an out-of-plane compression simulation where all units were bi-stable and required the same amount of out-of-plane displacement ($\Delta d$) to go from the first to the second stable configuration given by (See the Supporting Information for more details about the design of these eight units)

$$\Delta d = 2H \times \frac{\sqrt{B^2 - 4AC}}{A}, \tag{4}$$

where $H$ is defined in **Figure 1 (A)** and $A$, $B$ and $C$ can be determined using **Equation 2**. The strain energy and out-of-plane load vs. out-of-plane displacement of these eight units are plotted in **Figure 2(C)** and **(D)**. The overall shape of the energy curves in **Figure 2(C)** does not change for different values of $n$ and all eight units retain the zero-energy level at both zero and $10.2\ mm$ displacements. However, the point corresponding to maximum energy for these curves depend on the $n$ value. The three-pointed and six-pointed star-shaped units have the smallest and largest strain energies among the eight simulated units, respectively. This is confirmed in the load-displacement curve plotted in **Figure 2(D).** Furthermore, **Figure 2(E)** shows the distribution of strain energy density of three-pointed, six-pointed and ten-pointed star-shaped units at two stable and one maximum energy configurations. The dark blue color in two stable configurations confirms the zero amount of stored energy and bi-stability of these units. The distribution of strain energy is not uniform and it is higher near the creases which reveals the necessity of using brass pins in that area which has higher strength than the 3D printed material.

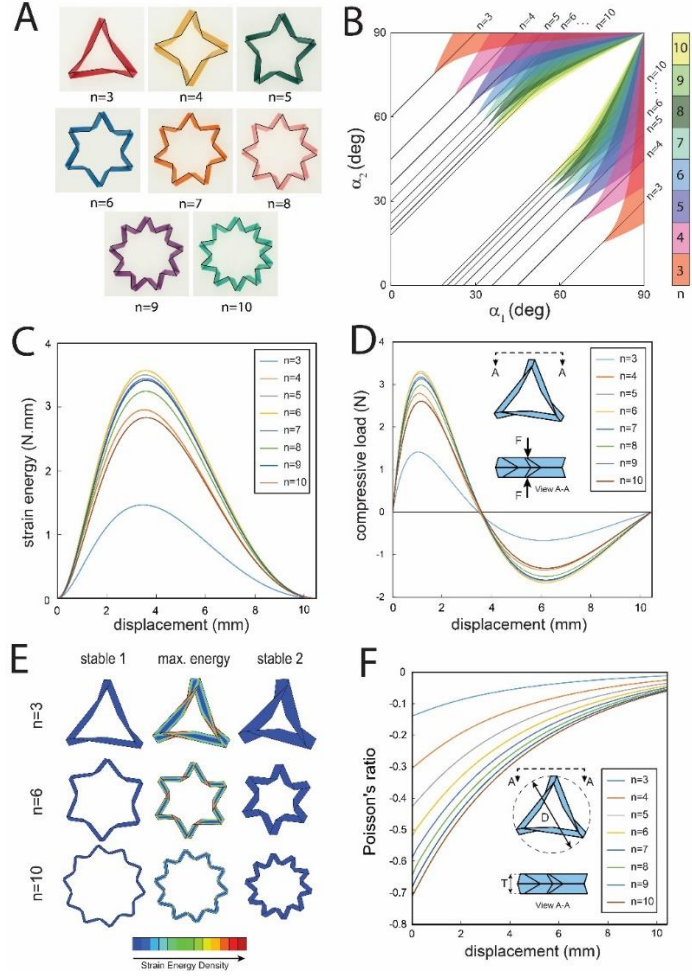

**Figure 2.** A) $n$-pointed ($3 \leq n \leq 10$) star-shaped units formed by folding origami strings with $n$ number of identical Miura-ori. B) Stability of $n$-pointed star-shaped units. The colored regions correspond to units with bi-stable and semi-bi-stable behaviors. C) Numerically obtained elastic strain energy versus applied out-of-plane compressive displacement for star-shaped units. D) Numerically determined force-displacement response of star-shaped units discussed in part C under out-of-plane compression. E) Strain energy density distribution for three star units with $n = 3$, $n = 6$ and $n = 10$ at two zero-energy (stable) and one maximum energy configurations. F) Poisson's ratio of the star-shaped units discussed in part C as a function of out-of-plane displacement.

Another interesting property in the bi-stable star-shaped units is the auxeticity (or negative Poisson's ratio) which can add a wide range of functionality to the structure such as tunable bandgap[24] and tunable shape[25]. The second stable configuration is attained through an in-plane contraction from the first stable configuration so that the resultant cross-sectional area is less than the initial area. The poison's ratio for unit is defined as $dD/dT$, where $D$ and $T$ are diameter of the circumcircle and height of the star-shaped unit, respectively, and shown in the schematic of **Figure 2(F)**. The Poisson's ratios of eight units discussed earlier ($3 \leq n \leq 10$) are calculated numerically using FE simulation as a function of displacement and results are presented in **Figure 2(F)**. All units have negative Poisson's ratio with decreasing absolute value as they deform toward the second stable configuration. Also, greater values of $n$ result in a larger absolute value of Poisson's ratio with $n = 10$ exhibiting the most intense auxetic behavior among the simulated units.

The unique properties of the proposed star-shaped units such as programming instability by changing the $\alpha_1$ and $\alpha_2$ angles, auxeticity and rotationally symmetric geometry make them a promising candidate to be used as building blocks of a cellular metamaterial (synthetic materials with nontraditional and extreme properties). The lattice of this type of cellular metamaterial could be made by tilling star-shaped units in 2D or 3D spaces. **Figures 3(A)** and **(B)** show two examples of such metamaterials made from four-pointed and eight-pointed star-shaped units. The cellular metamaterial shown in **Figure 3(A)** is consisted of 27 units with $\alpha_1 = 78.9°$, $\alpha_2 = 35.8°$, $a = 20\ mm$ and $H = 12\ mm$. Individual unit cells have two stable configurations at $\theta_1 = 18°$ and $\theta_2 = 62°$, determined from **Equation 3**, and switching between these two stable points causes reconfiguration of the unit cells. This reconfigurability gets transferred to the entire structure and creates two corresponding stable configurations at $\theta_1 = 18°$ and $\theta_2 = 62°$ in the lattice, see **Figure 3(A)** and the Supporting Movie for more details on this structure. This type of topologically dictated extreme behavior is hallmark of metamaterials [26-28]. Similarly, another metamaterial made by tiling four eight-pointed star-shaped units with $\alpha_1 = 72°$, $\alpha_2 = 51°$, $a = 20\ mm$ and $H = 15\ mm$ is shown in **Figure 3(B)**. Both metamaterial and individual cells retain two stable configurations at $\theta_1 = 18°$ and $\theta_2 = 60°$. The top views of metamaterial at two stable configurations are shown in **Figure 3(B)**. To characterize the reconfigurability of these metamaterials, we study the variations of cross-sectional area and internal volume as factors indicating the intensity of reconfiguration in two-dimensions and three-dimensions, respectively. The cross-sectional area is the area of the smallest rectangle which can cover the top or bottom surface of the metamaterial and the internal volume is the volume of the smallest cuboid which fits the metamaterial interior. **Figure 3(C)** and **(D)** show the variation of the cross-sectional area and the internal volume of metamaterials presented in **Figure 3(A)** and **(B)** as a function of out-of-plane displacement. The auxeticity of individual unit cells causes a decreasing rate in the area and volume such that the second stable configuration occupies smaller cross-sectional area and internal volume than the first stable configuration. As shown in the schematics and curves of **Figure 3(C)** and **(D)**, the cross-sectional area and internal volume of metamaterial with eight-pointed stars decrease at a higher rate than the metamaterial with four-pointed stars. This reveals the effect of unit cells type on the reconfigurability of the metamaterial. Another factor affecting the reconfigurability of the metamaterial is the $\alpha_1$ and $\alpha_2$ angles which define the properties of underlying star-shaped unit cells. The metamaterials are reconfigurable only in certain range of $\alpha_1$ and $\alpha_2$ as discussed earlier in the context of **Figure 1(C)**. These ranges are shown in **Figures 3(E)** and **(F)** for two previously introduced metamaterials and quantified in terms of shrinkage ratio is defined as the positive ratio of change in dimension $L$ (shown in subfigure) to the out-of-plane displacement between the first and second stable configurations. The color contours of **Figures 3(E)** and **(F)** show the variation of shrinkage ratio with respect to $\alpha_1$ and $\alpha_2$ for the metamaterials shown in **Figure 3(A)** and **(B)**, respectively. As $\alpha_1$ and $\alpha_2$ go toward 90°, shrinkage ratios go to zero and result in a structure which has two similar stable configurations but decreasing the values of $\alpha_1$ and $\alpha_2$, while they are still inside the colored area, causes more distinct stable configurations and maximize the reconfigurability of the structure.

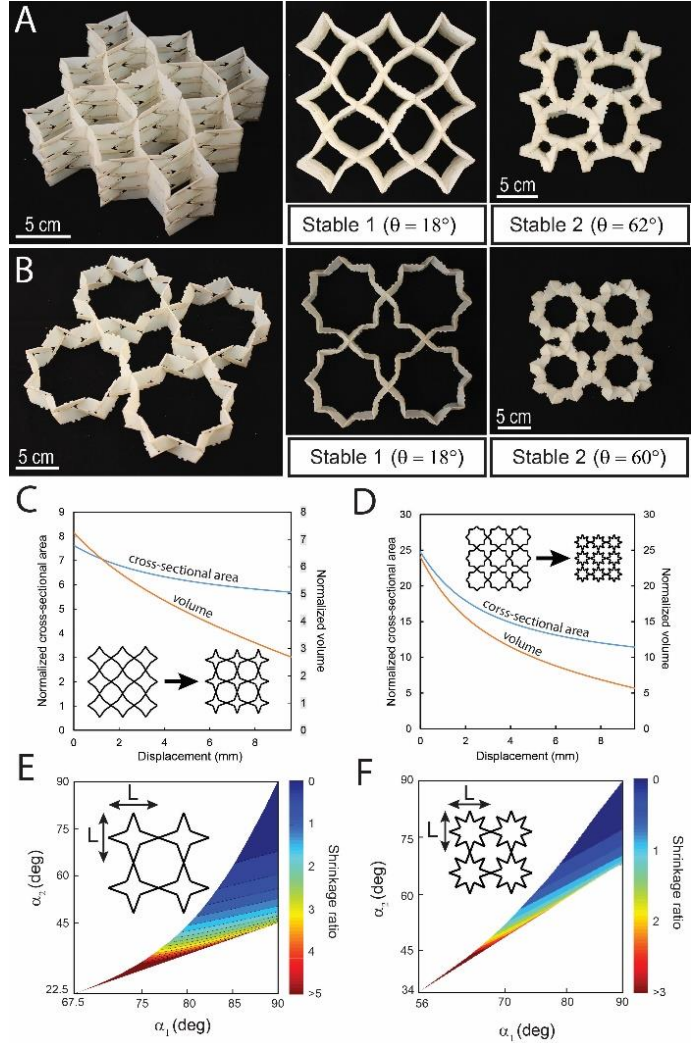

**Figure 3.** A) Isometric view of a 3 × 3 × 3 cellular structure with four-pointed star-shaped unit cells. Top view of two stable configurations are shown on the right. B) A 2 × 2 × 1 cellular structure with eight-pointed star unit cells. Top view of two zero-energy configurations (unfolded and folded) are shown on the right. C, D) Normalized cross-sectional area and normalized volume vs. out-of-plane displacement for two structures presented in part A and B. E, F) The shrinkage ratio of the lattices shown in parts A and B as a function of $\alpha_1$ and $\alpha_2$ angles.

The introduced star-shaped units can also be positioned in other ways to tailor the behavior of the metamaterial by introducing anisotropy (different response along different loading axes) and multi-stability. **Figure 4(A)** shows this novel metamaterial which is constructed from six four-pointed star-shaped units arranged to form a cuboid. **Figure 4(A)** also shows three configurations of the metamaterial, achieved by harnessing bi-stability of star units in each direction. The resulting metamaterial has zero effective Poisson's ratio in three orthogonal directions and multiple stable configurations that enable controlled reconfiguration in each $x$, $y$, and $z$ directions independent of other two orthogonal directions. The proposed metamaterial can be scaled in three directions and to any desired size by adding extra unit cells. Increasing the size of metamaterials significantly increases the number of design parameters,

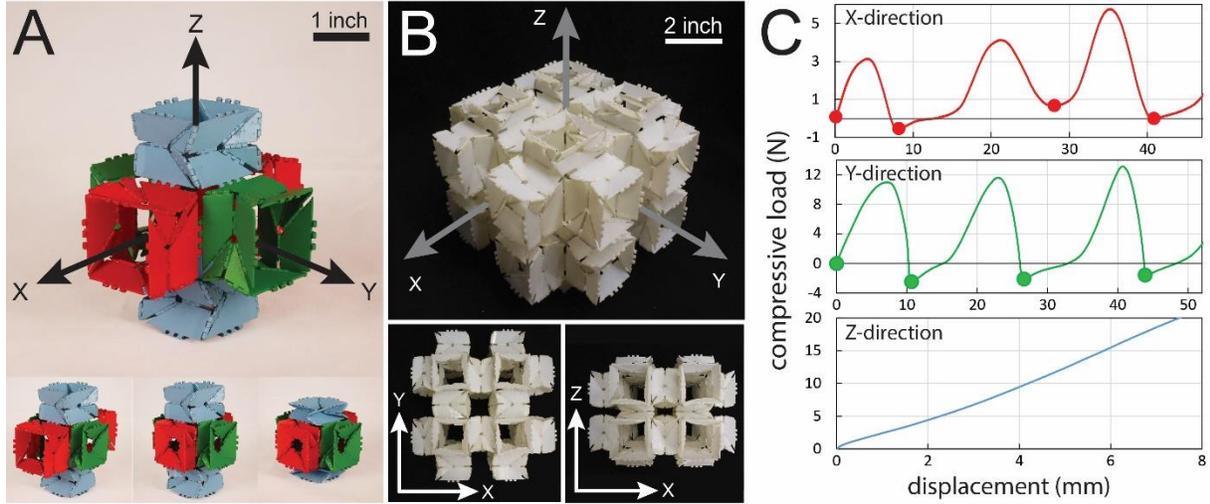

**Figure 4.** A) A cubic metamaterial which has multiple stable configurations as shown in the figure and constructed from six four-pointed star-shaped units positioned along three orthogonal directions. The metamaterial is anisotropic and can be programmed to have a desired response in each orthogonal direction by manipulating $\alpha_1$ and $\alpha_2$ angles of each star unit. B) Scaling the metamaterial by packing eight of them in a $2 \times 2 \times 2$ network. The resultant structure is multi-stable, anisotropic and has 15 independent design parameters in each orthogonal direction. C) Compressive load vs. displacement obtained from horizontal testing in $X, Y$, and $Z$ directions. The $2 \times 2 \times 2$ metamaterial is programmed to have stiffening multi-stability, monotonic multi-stability, and single-stability in compression along the $X, Y$, and $Z$ directions

resulting in a wide range of properties. This terminology is congruent with the results presented for star units where their properties are function of $\alpha_1$ and $\alpha_2$ characteristics angles, and these angles can be changed to tune the properties of the 3D metamaterial. Scalability of the proposed 3D metamaterial improves the design possibilities and response tunability. **Figure 4(B)** shows a metamaterial with 12 star-shaped units in each direction (36 totally) connected to neighboring units using revolute hinges. This metamaterial has 15 independent design parameters in each orthogonal direction, which can be used for programming the mechanical behavior of metamaterial in that direction. In general, having $n_1, n_2 \text{ and } n_3$ star-shaped units in series along the orthogonal directions provides $3(n_1 n_2 n_3) - 2(n_1 n_2 + n_2 n_3 + n_1 n_3) + 2(n_1 + n_2 + n_3)$ independent design parameters. The metamaterial shown in **Figure 4(B)** is programmed to have three different behaviors of; 1) stiffening multi-stability in which the required force for transition between consecutive stable configurations increases gradually, 2) monotonic multi-stability with constant required force for transition between consecutive stable configurations, and 3) single-stability with only one stable configuration in compression along the $x$, $y$, and $z$ directions. To do so, we have placed three different types of bi-stable star-shaped units along the $X$ axis, one type of bi-stable star-shaped unit along the $Y$ axis, and one type of single-stable star-shaped unit along the $Z$ axis. The horizontal compression test has been performed using an ADMET testing machine along all three orthogonal directions of the metamaterials and the obtained load-displacement results are shown in **Figure 4(C)**. The metamaterial exhibits multiple local minima along the $X$ direction (markers on the curve) in which removing force will result in the stable configuration. Also, as expected, an increasing peak values can be observed ($3.0, 4.1, 5.8\ N$). A similar response can be obtained from the compression testing along the $Y$ direction while the peak values are identical and equal to $11.8\ N$ (9% tolerance due to the error in fabrication process). The compression test in the Z direction shows an almost linear response with no stable configuration other than the initial point.

In conclusion, we introduced a novel family of origami-based structure fabricated by the folding a Miura-ori string. Our analytical investigation showed that by manipulating values of $\alpha_1, \alpha_2$ angles and number of Miura-ori ($n$) in the string one can create three distinct stability regimes - single-stability, bi-stability and semi-bi-stability. The experimental out-of-plane compression test on 3D printed prototypes as well as the FE simulation for different unit geometries confirm the analytical results and reveal the potential of star-shaped units to serve as building blocks of a metamaterial with programmable stability, reconfigurability, and anisotropy. This type of inherently lightweight and readily manufactured structure can have potentially transformative impact on a number of modern high performance industrial, medical, military and aerospace systems.

**Acknowledgement**

This work is supported by the United States National Science Foundation, Division of Civil, Mechanical, and Manufacturing Innovation, Grant No.1634560.

# Supplementary Information

# Origami-inspired Cellular Metamaterial with Anisotropic Multi-stability

Soroush Kamrava,[a] Ranajay Ghosh,[b] Zhihao Wang,[a] and Ashkan Vaziri [a,*]

[a] Department of Mechanical and Industrial Engineering,
Northeastern University, Boston, MA 02115, USA
[b] Department of Mechanical and Aerospace Engineering
University of Central Florida, Orlando, FL 32816, USA

*Corresponding Author: vaziri@coe.neu.edu

The configurations for the three different stability behaviors: single-stability, bi-stability and semi-bi-stability has been found through two criteria in the manuscript. The first criterion is $\eta_1 + \eta_2 = \pi(2 - 2/n)$ where $\eta_1$ and $\eta_2$ are interior angles of the star and shown in the **Figure 1(B)** as well as the **Figure 1s**. The second criterion is $\sum_{i=1}^{n} \vec{l_i} = 0$ representing the sum of middle crease vectors equating to zero.

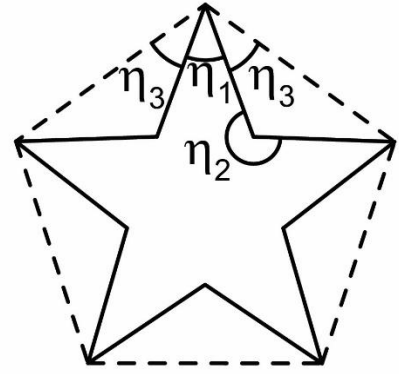

**Figure 1s.** Connecting sharp ends in the 5-pointed stars to form a convex polygon.

For deriving the equation for the first criterion, we sketch a $n$-pointed regular star (here $n = 5$) and connect the sharp ends using dashed lines, as shown in **Figure 1S**. These dashed lines form a convex five-sided polygon. The sum of its interior angles is given by $(n - 2)\pi$. Using angles $\eta_1$, $\eta_2$ and $\eta_3$ shown in the **Figure 1s**, we can rewrite the sum of interior angles in form of the following equation,

$$n \times (\eta_1 + 2\eta_3) = (n - 2) \times \pi. \tag{1s}$$

Also, interior angles of $\eta_3$, $\eta_3$ and $2\pi - \eta_2$ form a triangles and add up to $\pi$,

$$2\eta_3 + 2\pi - \eta_2 = \pi \tag{2s}$$

Determining $\eta_3$ from **Equation 2s** and plugging into the **Equation 1s** results in the following relationship between $\eta_1$ and $\eta_2$ (interior angles) for any $n$-pointed axisymmetric star

$$\eta_1 + \eta_2 = \pi(2 - 2/n), \tag{S3}$$

which is the first criterion used in the paper.

Eight different bi-stable samples in form of $n$-pointed star-shaped units ($n = 3, 4 \ldots, 10$) has been shown and studied in the **Figure 2**. These bi-stable units should be designed to require identical out-of-plane displacement for the transition from the first stable configuration to the second one. This similarity between the units enables us to consistently compare the bi-stability and geometrical behavior of units under the out-of-plane compression. This similarity is guaranteed by having same $\theta_1$ and $\theta_2$ angles, which represent the folding level of the first and second stable configurations among all eight units. Table 1s shows the values of $\alpha_1$ and $\alpha_2$ and the resulting $\theta_1$ and $\theta_2$ stable angles for all eight units.

**Table 1s.** Geometrical parameters and the stable configurations of eight samples shown in Figure 2.

| $n$ (number of end points in the star-shaped unit) | $\alpha_1$ (degree) | $\alpha_2$ (degree) | $\theta_1$ (degree) | $\theta_2$ (degree) |
|---|---|---|---|---|
| 3 | 80 | 20.5 | 30.11 | 69.9 |
| 4 | 74.7 | 30.3 | 29.94 | 69.57 |
| 5 | 71.3 | 35.8 | 29.71 | 71.24 |
| 6 | 68.9 | 39.4 | 29.94 | 70.25 |
| 7 | 67.2 | 41.9 | 29.67 | 71.15 |
| 8 | 65.8 | 43.7 | 29.94 | 70.42 |
| 9 | 64.7 | 45.1 | 30.22 | 69.36 |
| 10 | 63.9 | 46.2 | 29.70 | 71.48 |

As shown in the Table 1s, values of $\theta_1$ and $\theta_2$ are approximately equal (2 % error) which enable us to fairly compare different aspects of these eight units.

*Supplementary movie is available upon request.